\documentclass[a4paper, 11pt, oneside]{article}
\pdfoutput=1
\usepackage{jheppub}
\usepackage{amssymb}
\usepackage{epsfig}
\usepackage{graphicx}
\usepackage{subfigure}
\usepackage{dcolumn}
\usepackage{bm}
\usepackage[usenames ,dvipsnames]{xcolor}
\usepackage{slashed}

\begin{document}

\title{CP Violating $hW^+W^-$ Coupling in the Standard Model and Beyond }

\author[a,b]{
  Da~Huang
}
\emailAdd{dahuang@ua.pt}
\affiliation[a]{Departamento de F\'{\i}sica da Universidade de Aveiro and CIDMA, \\
 Campus de Santiago, 3810-183 Aveiro, Portugal.}
\affiliation[b]{National Astronomical Observatories, Chinese Academy of Sciences, Beijing, 100012, China}

\author[a]{
  Ant\'{o}nio~P.~Morais
}
\emailAdd{aapmorais@ua.pt}

\author[c,d]{Rui~Santos}
\emailAdd{rasantos@fc.ul.pt}
\affiliation[c]{Centro de F\'{\i}sica Te\'{o}rica e Computacional,
   Faculdade de Ci\^{e}ncias,\\
   Universidade de Lisboa, Campo Grande, Edif\'{\i}cio C8
  1749-016 Lisboa, Portugal.}
\affiliation[d]{ISEL - Instituto Superior de Engenharia de Lisboa,\\
  Instituto Polit\'ecnico de Lisboa
 1959-007 Lisboa, Portugal.}

\date{\today}
\abstract{
Inspired by the recent development in determining the property of the observed Higgs boson, we explore the $CP$-violating (CPV) $- c_{\rm CPV} h W^{+\, \mu\nu}\tilde{W}^{-}_{\mu\nu}/v$ coupling in the Standard Model (SM) and beyond, where $W^{\pm \, \mu \nu}$ and $\tilde{W}^{\pm\,\mu\nu}$ denote the $W$-boson field strength and its dual. To begin with, we show that the leading-order SM contribution to this CPV vertex appears at two-loop level. By summing over the quark flavor indices in the two loop integrals analytically, we can estimate the order of the corresponding Wilson coefficient to be $c^{\rm SM}_{\rm CPV} \sim {\cal O}(10^{-23})$, which is obviously too small to be probed at the LHC and planned future colliders. Then we investigate this CPV $hW^+ W^-$ interaction in two 
Beyond the Standard Model benchmark models: the left-right model and the complex 2-Higgs doublet model (C2HDM). Unlike what happens for the SM, the dominant contributions in both models 
arise at the one-loop level, and the corresponding Wilson coefficient can be as large as of ${\cal O}(10^{-9})$ in the former model and of ${\cal O}(10^{-3})$ for the latter. In light of such a large CPV effect in the $hW^+W^-$ coupling, we also give the formulae for the leading one-loop contribution to the related CPV $hZZ$ effective operator in the complex 2-Higgs doublet model. The order of magnitude of the Wilson coefficients in the C2HDM may be within reach of the high-luminosity LHC or planned future colliders. 
}

\maketitle

\section{Introduction}
\label{s1}
The discovery of the Higgs particle at the Large Hadron Collider (LHC) has finally completed the Standard Model (SM)~\cite{Aad:2012tfa,Chatrchyan:2012ufa,Chatrchyan:2013lba} of particle
physics. However, we still have the task of determining to what extent is the observed particle the genuine SM Higgs boson or if it is rather a scalar particle from a model
beyond the SM (BSM) with an extended scalar sector.
One can explore this question by carefully examining the properties of the discovered Higgs boson, $h$, such as its spin, its $CP$ properties and its couplings to the SM gauge bosons and fermions. 
Until now, all experimental results are consistent with the SM predictions. Nevertheless, with the luminosity increase during the next LHC stages, one still expects to pursue all
small deviations from the SM predictions. 
The anomalous $CP$-violating (CPV) $hW+W^-$ couplings have been probed and constrained at the LHC by both the CMS~\cite{Khachatryan:2014kca, Khachatryan:2016tnr,Sirunyan:2019twz,Sirunyan:2019nbs} and ATLAS~\cite{Aad:2013xqa,Aad:2015mxa,Aad:2016nal} collaborations. 
Concretely, the relevant anomalous interactions between the SM Higgs $h$ and a pair of $W^\pm$ bosons can be represented by the following scattering amplitude~\cite{Gao:2010qx,Bolognesi:2012mm,Anderson:2013afp,Plehn:2001nj,Hankele:2006ma, Hagiwara:2009wt,DeRujula:2010ys,Ellis:2012xd,Artoisenet:2013puc,Greljo:2015sla}:
\begin{eqnarray}\label{amp0}
{\cal M} (hW^+ W^-) \sim a_1^{W^+ W^-} m_W^2 \epsilon_{W^+}^* \epsilon_{W^-}^* + a_3^{W^+ W^-} f^{* +}_{\mu\nu} \tilde{f}^{*-\,\mu\nu}\,,
\end{eqnarray}  
where $f^{i\,\mu\nu} \equiv \epsilon_{W^i}^{\mu} q^{\nu}_i -\epsilon_{W^i}^\nu q_i^{\mu}$ is the $W$ gauge boson field strength tensor and $\tilde{f}^{i}_{\mu\nu} \equiv \epsilon_{\mu\nu\rho\sigma}f^{i\,\rho\sigma}/2$ is its dual field strength; $i=\pm$ and $q_i$ is the momentum of the $W$ boson with charge $i$. The anomalous CP violating contribution comes from the second term in Eq.~(\ref{amp0}), but we have 
also included in the amplitude of Eq.~(\ref{amp0}) the first term which already exists in the SM at the tree level. The most recent measurement on the CPV $hW^+ W^-$ coupling $a^{W^+ W^-}_3$ is given by the CMS experiment in Ref.~\cite{Sirunyan:2019nbs}, which has constrained this coupling to be in the range $a_3/a_1 \in  [-0.81, 0.31]$ at the $95\%$ confidence level (C.L.). From the effective
field theory perspective we can also represent this bound on the CPV coupling in terms of the Wilson coefficients of the corresponding effective operators. Note that, in Eq.~(\ref{amp0}), the CPV $hW^+ W^-$ term in the amplitude can be parametrised by the following effective operator~\cite{Plehn:2001nj,Hankele:2006ma,Artoisenet:2013puc}:
\begin{eqnarray}\label{EffectiveOperator}
{\cal O}_{\rm CPV} = -\frac{c_{\rm CPV}}{v} h W^{+ \, \mu\nu} \tilde{W}^-_{\mu\nu}\,
\end{eqnarray}
where $\tilde{W}^+_{\mu\nu} \equiv \epsilon_{\mu\nu\rho\sigma} W^{\pm\, \rho\sigma}/2$ is the dual $W$-boson field strength,
while the $CP$-conserving part is induced by 
\begin{eqnarray}
{\cal O}_{\rm CPC} = \frac{c_{\rm CPC}}{v} m_W^2 h W^{+ \, \mu\nu} \tilde{W}^-_{\mu\nu}\,.
\end{eqnarray}
If we further assume that the Wilson coefficient of the $CP$-conserving operator is taken to be its SM value $c_{\rm CPC} = 2$, then the experimentally allowed range of the coefficient of the CPV operator is given by
\begin{eqnarray}\label{ExpBound}
c_{\rm CPV} = 2\times \frac{a^{W^+ W^-}_3}{a^{W^+ W^-}_1} \in [-1.62,\, 0.62]\quad \mbox{at 95$\%$ C.L.}\,.
\end{eqnarray}

The importance of testing the properties of the observed Higgs particle led us to compute the size the CPV effect in the $hW^+W^-$ interaction in the SM and in two BSM models. 
It is well-known that the CPV $hW^+W^-$ Wilson coefficient is extremely small in the SM. Thus, the observation of this CPV effect at the LHC would constitute an unambiguous signal of new physics beyond the SM. However, there is no reasonable order of magnitude estimate for the SM value in the literature and neither a rigorous calculation in extensions of the SM where the values approach the ones that
are predicted by the experimental collaborations for the future LHC runs. We will then start by providing a reasonable order estimate of the Wilson coefficient of the CPV $hW^+ W^-$ effective operator in the SM. 
The leading-order contribution in the SM appears at the two-loop level, and is thus greatly suppressed by both the loop factor and the Glashow-Iliopoulos-Maiani (GIM) mechanism. As a result, the induced CPV $hW^+ W^-$ coupling in the SM cannot be probed at the LHC and even at future colliders. After establishing the non-detectability in the SM, we then further explore the CPV $hW^+ W^-$  coupling in two BSM models: the left-right model~\cite{Pati:1974yy, Mohapatra:1974gc, Senjanovic:1975rk, Fritzsch:1975yn}, and the complex 2-Higgs doublet model (C2HDM)~\cite{Weinberg:1976hu}. Since the CPV $hW^+W^-$ operator is generated at the one-loop level, a significant enhancement of the CPV effect is expected. We will calculate the typical magnitudes of the corresponding Wilson coefficients in these two benchmark models under the present constraints.

The paper is organised as follows. We firstly estimate the order of the magnitude of the CPV $hW^+ W^-$ effect in the SM in Sec.~\ref{sec_SM}. Then we move to the predictions of the Wilson coefficients of the CPV $hW^+ W^-$ effective operator in both the left-right model and the C2HDM in Sec.~\ref{sec_LR} and Sec.~\ref{sec_C2HDM}, respectively. Finally, we conclude in Sec.~\ref{sec_Conclusion}. In Appendix~\ref{sec_hZZ}, we calculate the magnitude of the related CPV $hZZ$ vertex in the C2HDM, in the light of its potential measurement in the future experiments.

\section{$CP$-Violating $hW^+W^-$ Coupling in the Standard Model}\label{sec_SM}
\begin{figure}[!ht]
\centering
\includegraphics[width = 0.75 \linewidth]{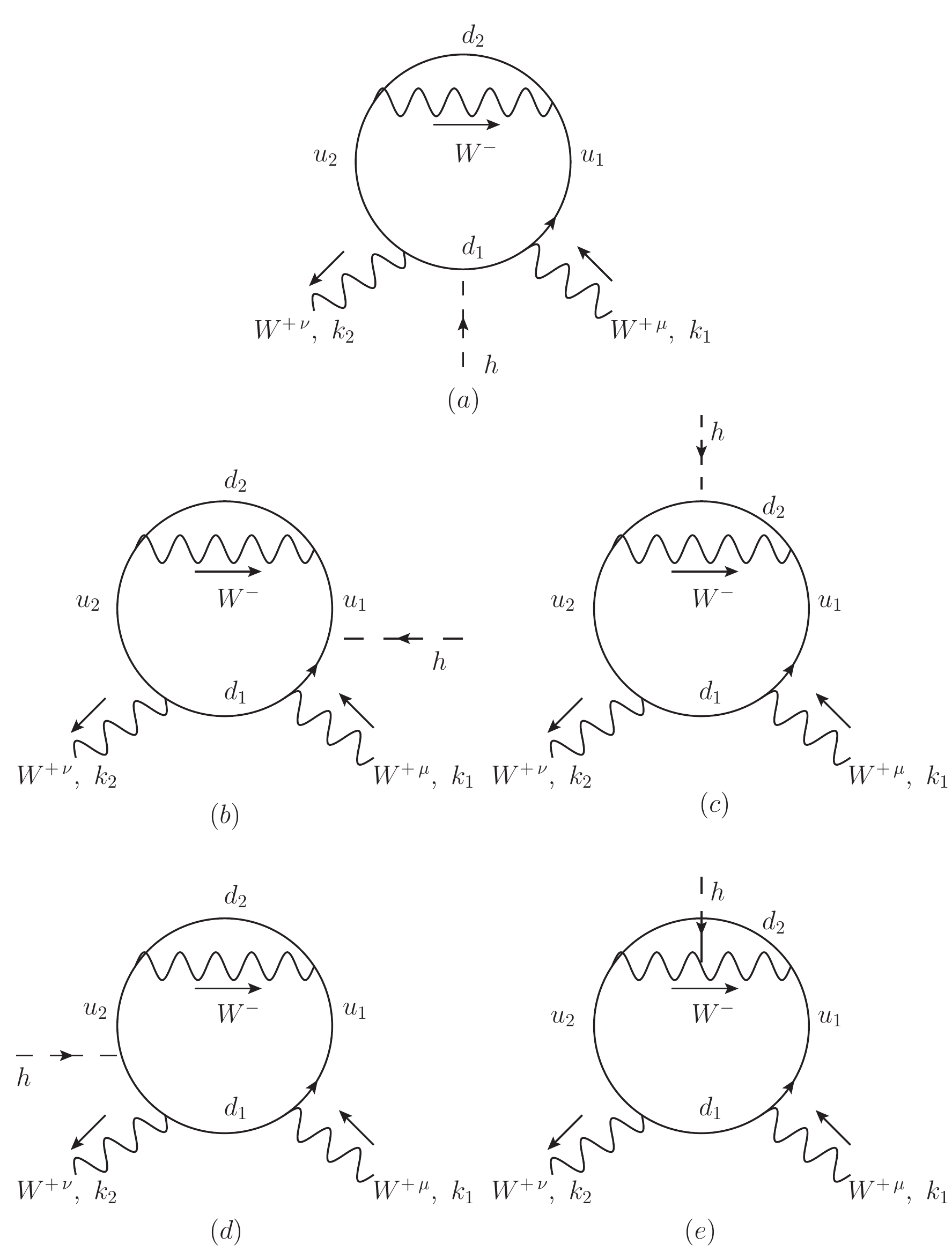}
\caption{Feynman diagrams leading to the CPV $hW^+W^-$ coupling in the SM.}\label{FigSM}
\end{figure}

It is well-known that in the SM all CPV effects arise from the CKM phase $\delta$, so that the corresponding Wilson coefficients should be proportional to the Jarlskog invariant $J={\rm Im}(V_{ud}V_{cd}^* V_{cs} V_{cd}^*) = c_{12} c_{23} c_{13}^2 s_{12} s_{23} s_{13} s_{\delta} = 3.00\times 10^{-5}$~\cite{Jarlskog:1985ht,Jarlskog:1985cw,Wu:1985ea,Tanabashi:2018oca}, where $s_{ij} \equiv {\rm sin}\theta_{ij}$ and $c_{ij} \equiv {\rm cos} \theta_{ij}$ with $\theta_{ij}$ representing the mixing angles between $i$ and $j$ quark families in the standard parametrization of the CKM matrix. An important consequence of this fact is that the CPV $hW^+ W^-$ vertex cannot be generated at tree or one-loop level, since there are not enough CKM matrix element insertions in the corresponding Feynman diagrams. Therefore, the CPV $hW^+W^-$ can only firstly appear at the two-loop order, with the corresponding five classes of Feynman diagrams shown in Fig.~\ref{FigSM}. 
The task of this section is to compute the analytic expressions of these five diagrams and extract the CPV contribution to the $hW^+ W^-$ coupling.
 
Let us begin our discussion by showing that the Feynman diagram (a) cannot contribute to any CPV effect. The amplitude of (a) can be written as follows:
\begin{eqnarray}\label{ampA}
i{\cal M}_{(a)} & = & (-1)N_c \int_{l_1} \int _{l_2}{\rm Tr}\Bigg[\left(-\frac{ig}{\sqrt{2}} V_{u_1 d_1}\gamma^\mu P_L\right)\frac{i}{\slashed{l}_1-\slashed{k}_1 -m_{d_1}} \left( -\frac{iy_{d_1}}{\sqrt{2}} \right) \frac{i}{\slashed{l}_1-\slashed{k}_2 -m_{d_1}}  \nonumber\\
&&  \times\left(-\frac{ig}{\sqrt{2}} V^*_{u_2 d_1}\gamma^\nu P_L\right) \frac{i}{\slashed{l}_1 -m_{u_2}} \left(-\frac{ig}{\sqrt{2}} V_{u_2 d_2}\gamma^\sigma P_L\right) \frac{i}{\slashed{l}_1 + \slashed{l}_2 -m_{d_2}}  \nonumber\\
&&  \times \left(-\frac{ig}{\sqrt{2}} V^*_{u_1 d_2}\gamma^\rho P_L\right) \frac{i}{\slashed{l}_1 - m_{u_1}} \Bigg]\frac{-i(g_{\rho \sigma} - l_{2\,\rho}l_{2\,\sigma}/m_W^2)}{l_2^2 - m_W^2}\nonumber\\
&=& iN_c \left(\frac{g}{\sqrt{2}}\right)^4 \frac{m_{d_1}^2}{v} \left(V_{u_1 d_1} V_{u_2 d_1}^* V_{u_2 d_2} V_{u_1 d_2}^* \right) \int_{l_1} \int_{l_2} \left(\frac{g_{\rho \sigma}-l_{2\,\rho} l_{2\,\sigma}/m_W^2}{l_2^2 -m_W^2}\right) \nonumber\\
&& \times \frac{{\rm Tr}[\gamma^\mu (2\slashed{l}_1-\slashed{k}_1 - \slashed{k}_2)\gamma^\nu \slashed{l}_1 \gamma^\sigma (\slashed{l}_1 + \slashed{l}_2)\gamma^\rho \slashed{l}_1 P_R]}{[(l_1 - k_1)^2 - m_{d_1}^2][(l_1-k_2)^2-m_{d_1}^2](l_1^2 - m_{u_2}^2)[(l_1+l_2)^2-m_{d_2}^2](l_1^2 - m_{u_1}^2)}\,,
\end{eqnarray}
where we have used the definition of the SM fermion mass $m_{f} = y_f v/\sqrt{2}$ with $y_f$ and $v$ the SM fermion Yukawa coupling and the SM Higgs vacuum expectation value (VEV), respectively. Here $N_c=3$ denotes the number of quark colors. Since we are only interested in the CPV part of the above amplitude, we focus on the imaginary part of the combination of CKM matrix elements ${\Phi}_{u_1 u_2}^{d_1 d_2} = {\rm Im}(V_{u_1 d_1} V_{u_2 d_1}^* V_{u_2 d_2} V_{u_1 d_2}^*)$, which has the following properties~\cite{Jarlskog:1985ht,Jarlskog:1985cw,Wu:1985ea,Dunietz:1987yt,Dunietz:1985uy}: (i) $\Phi_{u_1 u_2}^{d_1 d_2}$ is antisymmetric by interchanging the up- or down-quark indices: $\Phi_{u_1 u_2}^{d_1 d_2} = - \Phi_{u_2 u_1}^{d_1 d_2} = -\Phi_{u_1 u_2}^{d_2 d_1}$; (ii) the imaginary part should vanish when two up- or down-quark indices are identical, {\it i.e.}, $\Phi_{u_1 u_1}^{d_1 d_2} = \Phi_{u_1 u_2}^{d_1 d_1} =0$. Note that the second property guarantees that the two up- and down-quark flavor indices in the non-zero amplitude ${\cal M}_{(a)}$ should be different. Moreover, the two-loop integral in the second equality of Eq.~(\ref{ampA}) is symmetric under the interchange of the up-type quark indices $u_1 \leftrightarrow u_2$, which is inherited from their mass dependence. Thus, due to the antisymmetric factor $\Phi_{u_1 u_2}^{d_1 d_2}$, the final amplitude is antisymmetric under the swap of $u_1 \leftrightarrow u_2$. In order to obtain the total amplitude, we should sum up all of the flavor indices, including the up- and down-type quarks, which results in a vanishing contribution to the CPV $hW^+ W^-$ vertex. Finally, still for the class (a) of Feynman diagrams, we can generate new CPV contributions by changing the up(down)-type quarks into their down(up)-type quark counterparts. However, by considering the antisymmetry between the down-type quark indices this time, we would obtain the same result. Consequently, we will not account for the Feynman diagrams (a) any longer in the discussion. 

The contribution to the CPV $h W^+ W^-$ amplitude for the remaining four Feynman diagrams is not zero. For the class (b) of Feynman diagrams, the corresponding amplitude is given by
\begin{eqnarray}\label{ampB0}
i {\cal M}_{(b)} &=& (-1)N_c \int_{l_1} \int_{l_2} {\rm Tr}\Bigg[\left(-\frac{ig}{\sqrt{2}}V_{u_1 d_1}\gamma^\mu P_L \right) \frac{i}{\slashed{l}_1 -m_{d_1}}\left(-\frac{ig}{\sqrt{2}} V_{u_2 d_1}^* \gamma^\nu P_L \right) \frac{i}{\slashed{l}_1 +\slashed{k}_2 - m_{u_2}} \nonumber\\
&& \times \left(-\frac{ig}{\sqrt{2}} V_{u_2 d_2} \gamma^\sigma P_L \right)\frac{i}{\slashed{l}_1 + \slashed{l}_2 + \slashed{k}_2 - m_{d_2}} \left(-\frac{ig}{\sqrt{2}} V_{u_1 d_2}^* \gamma^\rho P_L \right)\frac{i}{\slashed{l}_1+\slashed{k}_2 - m_{u_1}} \nonumber\\
&& \times \left(-\frac{i y_{u_1}}{\sqrt{2}}\right) \frac{i}{\slashed{l}_1+\slashed{k}_1 - m_{u_1}} \Bigg] \frac{-i\left(g_{\rho\sigma} - l_{2\,\rho} l_{2\,\sigma}/m_W^2\right)}{l_2^2 - m_W^2} \nonumber\\
&=& i N_c \left(\frac{g}{\sqrt{2}}\right)^4 \frac{m_{u_1}^2}{v} \left(V_{u_1 d_1} V_{u_2 d_1}^* V_{u_2 d_2} V_{u_1 d_2}^* \right) \int_{l_1} \int_{l_2} \left(\frac{g_{\rho \sigma}-l_{2\,\rho} l_{2\,\sigma}/m_W^2}{l_2^2 -m_W^2}\right) \nonumber\\
&& \times \frac{{\rm Tr}[\gamma^\mu \slashed{l}_1 \gamma^\nu (\slashed{l}_1 + \slashed{k}_2)\gamma^\sigma (\slashed{l}_1+\slashed{l}_2 + \slashed{k}_2)\gamma^\rho (2\slashed{l}_1 + \slashed{k}_1 + \slashed{k}_2)P_R]}{(l_1^2-m_{d_1}^2)[(l_1+k_2)^2-m_{u_2}^2][(l_1+l_2+k_2)^2 -m_{d_2}^2]} \nonumber\\
&& \times \frac{1}{[(l_1+k_2)^2-m_{u_1}^2][(l_1+k_1)^2-m_{u_1}^2]}\,,
\end{eqnarray}
where we have used the $W$-boson propagator in the unitary gauge. 

We will now focus on the CPV part of the above amplitude, which should be proportional to $\Phi_{u_1 u_2}^{d_1 d_2} \equiv {\rm Im} \left(V_{u_1 d_1} V_{u_2 d_1}^* V_{u_2 d_2} V_{u_1 d_2}^* \right)$ as mentioned before. Also, we would like to sum over all the flavor indices to obtain the total contribution to the CPV effect. However, it is more illuminating to separate this summation into the following steps. The first step is to add up the two contributions with the interchange of $u_1 \leftrightarrow u_2$. Since $\Phi_{u_1 u_2}^{d_1 d_2}$ is antisymmetric under the exchange of $u_1 \leftrightarrow u_2$, the summation over these two terms is equivalent to antisymmetrize the up-type quark indices in the integral of Eq.~(\ref{ampB0}), yielding
\begin{eqnarray}
i{\cal M}_{(b)} &\sim& - \frac{N_c}{v} \left(\frac{g}{\sqrt{2}}\right)^4 \Phi_{u_1 u_2}^{d_1 d_2} \int_{l_1} \int_{l_2} \left(\frac{g_{\rho \sigma}-l_{2\,\rho} l_{2\,\sigma}/m_W^2}{l_2^2 -m_W^2}\right) \nonumber\\
&& \frac{{\rm Tr}[\gamma^\mu \slashed{l}_1 \gamma^\nu (\slashed{l}_1 + \slashed{k}_2)\gamma^\sigma (\slashed{l}_1+\slashed{l}_2 + \slashed{k}_2)\gamma^\rho (2\slashed{l}_1 + \slashed{k}_1 + \slashed{k}_2)P_R]}{(l_1^2-m_{d_1}^2)[(l_1+k_2)^2-m_{u_2}^2][(l_1+l_2+k_2)^2 -m_{d_2}^2][(l_1+k_2)^2-m_{u_1}^2]} \nonumber\\
&& \left[\frac{m_{u_1}^2}{(l_1+k_1)^2 -m_{u_1}^2} - \frac{m_{u_2}^2}{(l_1+k_1)^2 -m_{u_2}^2}\right]\nonumber\\
&=& -\frac{N_c}{v} \left(\frac{g}{\sqrt{2}}\right)^4 \Phi_{u_1 u_2}^{d_1 d_2} \int_{l_1} \int_{l_2} \left(\frac{g_{\rho \sigma}-l_{2\,\rho} l_{2\,\sigma}/m_W^2}{l_2^2 -m_W^2}\right) \nonumber\\
&& \frac{{\rm Tr}[\gamma^\mu \slashed{l}_1 \gamma^\nu (\slashed{l}_1 + \slashed{k}_2)\gamma^\sigma (\slashed{l}_1+\slashed{l}_2 + \slashed{k}_2)\gamma^\rho (2\slashed{l}_1 + \slashed{k}_1 + \slashed{k}_2)P_R]}{(l_1^2-m_{d_1}^2)[(l_1+k_2)^2-m_{u_2}^2][(l_1+l_2+k_2)^2 -m_{d_2}^2][(l_1+k_2)^2-m_{u_1}^2]} \nonumber\\
&& \times \frac{(m_{u_1}^2 - m_{u_2}^2)(l_1+k_1)^2}{[(l_1+k_1)^2 -m_{u_1}^2][(l_1+k_1)^2 - m_{u_2}]} 
\end{eqnarray}
where the symbol $\sim$ refers to the extraction of the CPV part of the amplitude. If we further take into account the antisymmetry between the indices $d_1$ and $d_2$ in $\Phi_{u_1 u_2}^{d_1 d_2}$, the summation over the terms with the interchange of the flavor indices $d_1$ and $d_2$ leads to
\begin{eqnarray}\label{ampB1}
i{\cal M}_{(b)} &\sim & -\frac{N_c}{v} \left(\frac{g}{\sqrt{2}}\right)^4 \Phi_{u_1 u_2}^{d_1 d_2} (m_{u_1}^2 - m_{u_2}^2) \int_{l_1} \int_{l_2} \left(\frac{g_{\rho \sigma}-l_{2\,\rho} l_{2\,\sigma}/m_W^2}{l_2^2 -m_W^2}\right) \nonumber\\
&& \frac{(l_1+k_1)^2 {\rm Tr}[\gamma^\mu \slashed{l}_1 \gamma^\nu (\slashed{l}_1 + \slashed{k}_2)\gamma^\sigma (\slashed{l}_1+\slashed{l}_2 + \slashed{k}_2)\gamma^\rho (2\slashed{l}_1 + \slashed{k}_1 + \slashed{k}_2)P_R]}{[(l_1+k_1)^2 -m_{u_1}^2][(l_1+k_1)^2 - m_{u_2}][(l_1+k_2)^2-m_{u_2}^2][(l_1+k_2)^2-m_{u_1}^2]} \nonumber\\
&& \times \left(\frac{1}{(l_1^2-m_{d_1}^2)(l_1+l_2+k_2)^2 -m_{d_2}^2]} - \frac{1}{(l_1^2-m_{d_2}^2)(l_1+l_2+k_2)^2 -m_{d_1}^2]} \right) \nonumber\\
&=& -\frac{N_c}{v} \left(\frac{g}{\sqrt{2}}\right)^4 \Phi_{u_1 u_2}^{d_1 d_2} (m_{u_1}^2 - m_{u_2}^2) \int_{l_1} \int_{l_2} \left(\frac{g_{\rho \sigma}-l_{2\,\rho} l_{2\,\sigma}/m_W^2}{l_2^2 -m_W^2}\right) \nonumber\\
&& \frac{(l_1+k_1)^2 {\rm Tr}[\gamma^\mu \slashed{l}_1 \gamma^\nu (\slashed{l}_1 + \slashed{k}_2)\gamma^\sigma (\slashed{l}_1+\slashed{l}_2 + \slashed{k}_2)\gamma^\rho (2\slashed{l}_1 + \slashed{k}_1 + \slashed{k}_2)P_R]}{[(l_1+k_1)^2 -m_{u_1}^2][(l_1+k_1)^2 - m_{u_2}][(l_1+k_2)^2-m_{u_2}^2][(l_1+k_2)^2-m_{u_1}^2]} \nonumber\\
&& \times \frac{(m_{d_1}^2 - m_{d_2}^2)[(l_1+l_2+k_2)^2-l_1^2]}{(l_1^2-m_{d_1}^2)(l_1^2 - m_{d_2}^2)[(l_1+l_2+k_2)^2 -m_{d_1}^2][(l_1+l_2+k_2)^2 -m_{d_2}^2]} \,.
\end{eqnarray}
Note that Eq.~(\ref{ampB1}) is valid only for a specific pair of up(down)-quark flavors, so our next task is to sum up all quark flavors. To begin with, we note that the factors $\Phi_{u_1 u_2}^{d_1 d_2}$ are all proportional to the Jarlskog parameter $J$ in the SM. In order to simplify our discussion we first fix the down-type quark flavors to be bottom, $b$ and strange, $s$, quarks, and sum over the corresponding up-type quark flavors. In this case, the relevant imaginary part of the CKM matrix elements combination has the following relation: $\Phi_{tc}^{bs} = -\Phi_{tu}^{bs} = \Phi_{cu}^{bs}=J$. Therefore, the summation over the up-type quark flavors leads to the following expression for the type-(b) Feynman diagrams
\begin{eqnarray}
i{\cal M}_{(b)} &\sim & -\frac{N_c J}{v} \left(\frac{g}{\sqrt{2}}\right)^4 (m_{b}^2 - m_{s}^2) \int_{l_1} \int_{l_2} \left(\frac{g_{\rho \sigma}-l_{2\,\rho} l_{2\,\sigma}/m_W^2}{l_2^2 -m_W^2}\right) \nonumber\\
&& \frac{(l_1+k_1)^2 [(l_1+l_2+k_2)^2-l_1^2]}{(l_1^2-m_{d_1}^2)(l_1^2 - m_{d_2}^2)[(l_1+l_2+k_2)^2 -m_{d_1}^2][(l_1+l_2+k_2)^2 -m_{d_2}^2]}
 \nonumber\\
&& \times  {\rm Tr}[\gamma^\mu \slashed{l}_1 \gamma^\nu (\slashed{l}_1 + \slashed{k}_2)\gamma^\sigma (\slashed{l}_1+\slashed{l}_2 + \slashed{k}_2)\gamma^\rho (2\slashed{l}_1 + \slashed{k}_1 + \slashed{k}_2)P_R] \nonumber\\
&& \times \Bigg\{\frac{(m_{t}^2 - m_{c}^2)} {[(l_1+k_1)^2 -m_{t}^2][(l_1+k_1)^2 - m_{c}][(l_1+k_2)^2-m_{t}^2][(l_1+k_2)^2-m_{c}^2]} \nonumber\\
&&\quad -  \frac{(m_{t}^2 - m_{u}^2)} {[(l_1+k_1)^2 -m_{t}^2][(l_1+k_1)^2 - m_{u}][(l_1+k_2)^2-m_{t}^2][(l_1+k_2)^2-m_{u}^2]}\nonumber\\
&& \quad  + \frac{(m_{c}^2 - m_{u}^2)} {[(l_1+k_1)^2 -m_{c}^2][(l_1+k_1)^2 - m_{u}][(l_1+k_2)^2-m_{c}^2][(l_1+k_2)^2-m_{u}^2]} \Bigg\} \nonumber\\
&=& -\frac{N_c J}{v} \left(\frac{g}{\sqrt{2}}\right)^4 (m_{b}^2 - m_{s}^2) \int_{l_1} \int_{l_2} \left(\frac{g_{\rho \sigma}-l_{2\,\rho} l_{2\,\sigma}/m_W^2}{l_2^2 -m_W^2}\right) \nonumber\\
&& \frac{(l_1+k_1)^2 [(l_1+l_2+k_2)^2-l_1^2]}{(l_1^2-m_{b}^2)(l_1^2 - m_{s}^2)[(l_1+l_2+k_2)^2 -m_{b}^2][(l_1+l_2+k_2)^2 -m_{s}^2]}
 \nonumber\\
&& \times  \frac{ \prod_{i>j} (m_{u_i}^2 - m_{u_j}^2) {\rm Tr}[\gamma^\mu \slashed{l}_1 \gamma^\nu (\slashed{l}_1 + \slashed{k}_2)\gamma^\sigma (\slashed{l}_1+\slashed{l}_2 + \slashed{k}_2)\gamma^\rho (2\slashed{l}_1 + \slashed{k}_1 + \slashed{k}_2)P_R]}{\prod_i [(l_1+k_1)^2 - m_{u_i}][(l_1+k_2)^2-m_{u_i}^2] } \,,
\end{eqnarray}
where the indices $i,j = 1,2,3$ denote different quark families. For the summation over the down-type quark flavors, a similar argument can give us the following expression for the total CPV amplitude of the Feynman diagrams of class (b)
\begin{eqnarray}\label{ampBm}
i{\cal M}_{(b)} &\sim & -\frac{N_c J}{v} \left(\frac{g}{\sqrt{2}}\right)^4  \int_{l_1} \int_{l_2} \left(\frac{g_{\rho \sigma}-l_{2\,\rho} l_{2\,\sigma}/m_W^2}{l_2^2 -m_W^2}\right) \nonumber\\
&& \times {\rm Tr}[\gamma^\mu \slashed{l}_1 \gamma^\nu (\slashed{l}_1 + \slashed{k}_2)\gamma^\sigma (\slashed{l}_1+\slashed{l}_2 + \slashed{k}_2)\gamma^\rho (2\slashed{l}_1 + \slashed{k}_1 + \slashed{k}_2)P_R] \nonumber\\
&& \times  \frac{ \prod_{i>j} (m_{u_i}^2 - m_{u_j}^2) (m_{d_i}^2-m_{d_j}^2) (l_1+k_1)^2 [(l_1+l_2+k_2)^2-l_1^2] }{\prod_i [(l_1+k_1)^2 - m_{u_i}][(l_1+k_2)^2-m_{u_i}^2](l_1^2-m_{d_i}^2)[(l_1+l_2+k_2)^2 -m_{d_i}^2] }\,.
\end{eqnarray}

Now we consider the other three classes of Feynman diagrams in Fig.~\ref{FigSM}.  We can write down the expressions for one specific flavor $(u_1, u_2; d_1, d_2)$ dependence for each class as follows
\begin{eqnarray}\label{ampC0}
i{\cal M}_{(c)} &=& (-1)N_c \int_{l_1} \int_{l_2} {\rm Tr}\Bigg[\left(-\frac{ig}{\sqrt{2}}V_{u_1 d_1}\gamma^\mu P_L \right) \frac{i}{\slashed{l}_1 -m_{d_1}}\left(-\frac{ig}{\sqrt{2}} V_{u_2 d_1}^* \gamma^\nu P_L \right) \frac{i}{\slashed{l}_1 +\slashed{k}_2 - m_{u_2}} \nonumber\\
&& \times \left(-\frac{ig}{\sqrt{2}} V_{u_2 d_2} \gamma^\sigma P_L \right)\frac{i}{\slashed{l}_1 + \slashed{l}_2 + \slashed{k}_2 - m_{d_2}} \left(-\frac{i y_{d_2}}{\sqrt{2}}\right) \frac{i}{\slashed{l}_1 + \slashed{l}_2 + \slashed{k}_1 - m_{d_2}} \nonumber\\
&& \times  \left(-\frac{ig}{\sqrt{2}} V_{u_1 d_2}^* \gamma^\rho P_L \right)\frac{i}{\slashed{l}_1+\slashed{k}_1 - m_{u_1}}  \Bigg] \frac{-i\left(g_{\rho\sigma} - l_{2\,\rho} l_{2\,\sigma}/m_W^2\right)}{l_2^2 - m_W^2} \nonumber\\
&=& i N_c \left(\frac{g}{\sqrt{2}}\right)^4 \frac{m_{d_2}^2}{v} \left(V_{u_1 d_1} V_{u_2 d_1}^* V_{u_2 d_2} V_{u_1 d_2}^* \right) \int_{l_1} \int_{l_2} \left(\frac{g_{\rho \sigma}-l_{2\,\rho} l_{2\,\sigma}/m_W^2}{l_2^2 -m_W^2}\right) \nonumber\\
&& \times\frac{{\rm Tr}[\gamma^\mu \slashed{l}_1 \gamma^\nu (\slashed{l}_1 + \slashed{k}_2)\gamma^\sigma (2\slashed{l}_1+2 \slashed{l}_2 + \slashed{k}_1 + \slashed{k}_2)\gamma^\rho (\slashed{l}_1 + \slashed{k}_1)P_R]}{(l_1^2-m_{d_1}^2)[(l_1+k_2)^2-m_{u_2}^2][(l_1+k_1)^2-m_{u_1}^2]} \nonumber\\
&& \times \frac{1}{[(l_1 + l_2 + k_2)^2-m_{d_2}^2][(l_1+l_2+k_1)^2-m_{d_2}^2]}\,,
\end{eqnarray}

\begin{eqnarray}\label{ampD0}
i{\cal M}_{(d)} &=& (-1)N_c \int_{l_1} \int_{l_2} {\rm Tr}\Bigg[\left(-\frac{ig}{\sqrt{2}}V_{u_1 d_1}\gamma^\mu P_L \right) \frac{i}{\slashed{l}_1 -m_{d_1}}\left(-\frac{ig}{\sqrt{2}} V_{u_2 d_1}^* \gamma^\nu P_L \right) \frac{i}{\slashed{l}_1 +\slashed{k}_2 - m_{u_2}} \nonumber\\
&& \times \left(-\frac{i y_{u_2}}{\sqrt{2}}\right) \frac{i}{\slashed{l}_1 +\slashed{k}_1 - m_{u_2}} \left(-\frac{ig}{\sqrt{2}} V_{u_2 d_2} \gamma^\sigma P_L \right) \frac{i}{\slashed{l}_1 + \slashed{l}_2 + \slashed{k}_1 - m_{d_2}} \nonumber\\
&& \times  \left(-\frac{ig}{\sqrt{2}} V_{u_1 d_2}^* \gamma^\rho P_L \right)\frac{i}{\slashed{l}_1+\slashed{k}_1 - m_{u_1}}  \Bigg] \frac{-i\left(g_{\rho\sigma} - l_{2\,\rho} l_{2\,\sigma}/m_W^2\right)}{l_2^2 - m_W^2} \nonumber\\
&=& i N_c \left(\frac{g}{\sqrt{2}}\right)^4 \frac{m_{u_2}^2}{v} \left(V_{u_1 d_1} V_{u_2 d_1}^* V_{u_2 d_2} V_{u_1 d_2}^* \right) \int_{l_1} \int_{l_2} \left(\frac{g_{\rho \sigma}-l_{2\,\rho} l_{2\,\sigma}/m_W^2}{l_2^2 -m_W^2}\right) \nonumber\\
&& \times\frac{{\rm Tr}[\gamma^\mu \slashed{l}_1 \gamma^\nu (2\slashed{l}_1 + \slashed{k}_1 + \slashed{k}_2) \gamma^\sigma (\slashed{l}_1+ \slashed{l}_2 + \slashed{k}_1)\gamma^\rho (\slashed{l}_1 + \slashed{k}_1)P_R]}{(l_1^2-m_{d_1}^2)[(l_1+l_2+k_1)^2-m_{d_2}^2][(l_1+k_1)^2-m_{u_1}^2]} \nonumber\\
&& \times \frac{1}{[(l_1 + k_2)^2-m_{u_2}^2][(l_1+k_1)^2-m_{u_2}^2]}\,,
\end{eqnarray}

\begin{eqnarray}\label{ampE0}
i{\cal M}_{(e)} &=& (-1)N_c \int_{l_1} \int_{l_2} {\rm Tr}\Bigg[\left(-\frac{ig}{\sqrt{2}}V_{u_1 d_1}\gamma^\mu P_L \right) \frac{i}{\slashed{l}_1 -m_{d_1}}\left(-\frac{ig}{\sqrt{2}} V_{u_2 d_1}^* \gamma^\nu P_L \right) \frac{i}{\slashed{l}_1 +\slashed{k}_2 - m_{u_2}} \nonumber\\
&& \times \left(-\frac{ig}{\sqrt{2}} V_{u_2 d_2} \gamma^\sigma P_L \right) \frac{i}{\slashed{l}_1 - \slashed{l}_2- m_{d_2}} \left(-\frac{ig}{\sqrt{2}} V_{u_1 d_2}^* \gamma^\rho P_L \right)\frac{i}{\slashed{l}_1+\slashed{k}_1 - m_{u_1}}  \Bigg] (i g m_W g^{\alpha\beta}) \nonumber\\
&& \times \frac{(-i)\left[g_{\sigma \beta} - (l_2 + k_2)_\sigma (l_2+k_2)_\beta /m_W^2\right]}{(l_2+k_2)^2-m_W^2} \frac{(-i)\left[g_{\rho \alpha} - (l_2 + k_1)_\rho (l_2+k_1)_\alpha /m_W^2\right]}{(l_2+k_1)^2-m_W^2} \nonumber\\
&=& i {N_c (g m_W)} \left(\frac{g}{\sqrt{2}}\right)^4 \left(V_{u_1 d_1} V_{u_2 d_1}^* V_{u_2 d_2} V_{u_1 d_2}^* \right) \int_{l_1} \int_{l_2} \nonumber\\
&& \times \frac{{\rm Tr}[\gamma^\mu \slashed{l}_1 \gamma^\nu (\slashed{l}_1 + \slashed{k}_2)\gamma^\sigma (\slashed{l}_1 - \slashed{l}_2)\gamma^\rho (\slashed{l}_1 + \slashed{k}_1)P_R]}{(l_1^2 - m_{d_1}^2)[(l_1 + k_2)^2 - m_{u_2}^2][(l_1-l_2)^2-m_{d_2}^2][(l_1+k_1)^2-m_{u_1}^2]}\nonumber\\
&& \times \frac{\left[g_{\sigma}^{~\alpha} - (l_2 + k_2)_\sigma (l_2+k_2)^\alpha /m_W^2\right] \left[g_{\rho \alpha} - (l_2 + k_1)_\rho (l_2+k_1)_\alpha /m_W^2\right]}{[(l_2+k_2)^2 -m_W^2][(l_2+k_1)^2-m_W^2]}\,.
\end{eqnarray}

Next we can sum up all of the flavor indices with the method used to treat the class (b) of Feynman diagrams. Since the procedure is almost the same, here we only list the final results of the CPV amplitudes for the remaining classes as
\begin{eqnarray}\label{ampCm}
i{\cal M}_{(c)} &\sim & - \frac{N_c J}{v} \left(\frac{g}{\sqrt{2}}\right)^4 \int_{l_1}\int_{l_2} \left(\frac{g_{\rho\sigma} - l_{2\,\rho}l_{2\,\sigma}/m_W^2}{l_2^2-m_W^2}\right) \nonumber\\
&& \times {\rm Tr}[\gamma^\mu \slashed{l}_1 \gamma^\nu (\slashed{l}_1 + \slashed{k}_2)\gamma^\sigma (2\slashed{l}_1+2 \slashed{l}_2 + \slashed{k}_1 + \slashed{k}_2)\gamma^\rho (\slashed{l}_1 + \slashed{k}_1)P_R] \nonumber\\
&& \times \frac{\prod_{i>j} (m_{u_i}^2-m_{u_j}^2) [(l_1+k_2)^2-(l_1+k_1)^2]}{\prod_i[(l_1+k_1)^2-m_{u_i}^2][(l_1+k_2)^2-m_{u_i}^2]}\nonumber\\
&& \times \frac{K_c\prod_{i>j}(m_{d_i}^2-m_{d_j}^2)}{\prod_i (l_1^2-m_{d_i}^2)[(l_1+l_2+k_1)^2-m_{d_i}^2][(l_1+l_2+k_2)^2-m_{d_i}^2]}\,,
\end{eqnarray}
where the factor $K_c$ is defined as
\begin{eqnarray}
K_c &\equiv & L_1^2 L_2^2 (L_1^2 L_2^2 -l_1^2 L_1^2 - l_1^2 L_2^2)+ l_1^2 L_1^2 L_2^2 (m_b^2+m_s^2 + m_d^2)\nonumber\\
&& - L_1^2 L_2^2(m_b^2 m_s^2 + m_s^2 m_d^2+ m_b^2 m_d^2)+ (L_1^2 + L_2^2 -l_1^2)m_b^2 m_s^2 m_d^2\,,
\end{eqnarray}
with 
\begin{eqnarray}
L_1\equiv l_1+l_2+k_1\,,\quad\quad L_2 \equiv l_1+l_2+k_2\,.
\end{eqnarray}

\begin{eqnarray}\label{ampDm}
i{\cal M}_{(d)} &\sim & \frac{N_c J}{v} \left(\frac{g}{\sqrt{2}}\right)^4 \int_{l_1} \int_{l_2} \left(\frac{g_{\rho \sigma}-l_{2\,\rho}l_{2\,\sigma}/m_W^2}{l_2^2 - m_W^2}\right) \nonumber\\
&& \times {\rm Tr}[\gamma^\mu \slashed{l}_1 \gamma^\nu (2\slashed{l}_1 + \slashed{k}_1 + \slashed{k}_2) \gamma^\sigma (\slashed{l}_1+ \slashed{l}_2 + \slashed{k}_1)\gamma^\rho (\slashed{l}_1 + \slashed{k}_1)P_R] \nonumber\\
&& \times \frac{\prod_{i>j} (m_{u_i}^2 - m_{u_j}^2) (m_{d_i}^2 - m_{d_j}^2)(l_1+k_2)^2[(l_1+l_2+k_1)^2-l_1^2]}{\prod_i [(l_1+k_1)^2 - m_{u_i}^2][(l_1+k_2)^2-m_{u_i}^2][l_1^2-m_{d_i}^2][(l_1+l_2+k_1)^2-m_{d_i}^2]}\,,
\end{eqnarray}

\begin{eqnarray}\label{ampEm}
i{\cal M}_{(e)} &\sim & - \frac{2 N_c J m_W^2}{v} \left(\frac{g}{\sqrt{2}}\right)^4 \int_{l_1} \int_{l_2} {\rm Tr}[\gamma^\mu \slashed{l}_1 \gamma^\nu (\slashed{l}_1 + \slashed{k}_2)\gamma^\sigma (\slashed{l}_1 - \slashed{l}_2)\gamma^\rho (\slashed{l}_1 + \slashed{k}_1)P_R] \nonumber\\
&& \times \frac{\prod_{i>j} (m_{u_i}^2 - m_{u_j}^2)(m_{d_i}^2 - m_{d_j}^2)[(l_1+k_2)^2 - (l_1+k_1)^2][(l_1 - l_2)^2-l_1^2]}{\prod_i [(l_1+k_1)^2-m_{u_i}^2][(l_1+k_2)^2-m_{u_i}^2](l_1^2-m_{d_i}^2)[(l_1 - l_2)^2 -m_{d_i}^2]} \nonumber\\
&& \times \frac{\left[g_{\sigma}^{~\alpha} - (l_2 + k_2)_\sigma (l_2+k_2)^\alpha /m_W^2\right] \left[g_{\rho \alpha} - (l_2 + k_1)_\rho (l_2+k_1)_\alpha /m_W^2\right]}{[(l_2+k_2)^2 -m_W^2][(l_2+k_1)^2-m_W^2]}\,.
\end{eqnarray} 

As mentioned in the discussion of Feynman diagrams of class (a), we can generate new contributions to the CPV $hW^+W^-$ amplitude by exchanging the role of up-type and down-type quarks in other diagrams of Fig.~\ref{FigSM}. The associated analytic formulae for these diagrams can be easily obtained by swapping the up-type and down-type quark notations of the same generation in Eqs.~(\ref{ampBm}), (\ref{ampCm}), (\ref{ampDm}) and (\ref{ampEm}).

From the expressions in Eqs.~(\ref{ampBm}), (\ref{ampCm}), (\ref{ampDm}) and (\ref{ampEm}), it is obvious that the CPV part of the total amplitude should be proportional to the following common factor:
\begin{equation}\label{CommFac}
\frac{N_c J }{v} \left(\frac{g}{\sqrt{2}}\right)^4 \prod_{i>j} (m_{u_i}^2 - m_{u_j}^2)(m_{d_i}^2 - m_{d_j}^2)\,.
\end{equation}
Note that this factor is actually dictated by the GIM mechanism~\cite{Glashow:1970gm} that works in the SM, since the $CP$-violation vanishes when any pair of up- or down-quark masses is the same. Nevertheless, here we have explicitly shown its origin by summing over the quark flavor indices in the two-loop Feynman integrals.  

We then proceed by noticing that the natural characteristic energy scale of two-loop integrals should be the $W$-boson mass $m_W$. Therefore, by applying this scale to balance the mass dimension of the loop integrals, the Wilson coefficient of the CPV effective operator in Eq.~(\ref{EffectiveOperator}) in the SM can be estimated to be
\begin{eqnarray}
|c_{\rm CPV}^{\rm SM}| \sim \frac{N_c J}{(16\pi^2)^2}\left(\frac{g}{\sqrt{2}}\right)^4 \frac{\prod_{i>j} (m_{u_i}^2 - m_{u_j}^2)(m_{d_i}^2 - m_{d_j}^2)}{m_W^{12}}\simeq 9.1\times 10^{-24} \sim {\cal O}(10^{-23})\,.
\end{eqnarray}
By comparing the current precision achieved experimentally for the anomalous $hW^+ W^-$ coupling in Eq.~(\ref{ExpBound}), the tiny SM prediction of this CPV effect cannot be observed under the present technology.

Finally, it is interesting to note that if the external Higgs boson is replaced by an external photon in Fig.~(\ref{FigSM}), the corresponding Feynman diagrams would induce the electric dipole moment of the $W$-boson~\cite{Pospelov:1991zt,Booth:1993af,Chang:1990fp}, another CPV quantity that has been widely studied in the literature. In the latter case, the two-loop contribution has been shown to vanish due to the Ward identity in QED that connects the $\bar{q}q^\prime \gamma$ vertex correction to those of quark $q,q^\prime$ masses at the one-loop level~\cite{Pospelov:1991zt}. However, we do not expect this cancellation would happen in the case of the CPV $hW^+W^-$ coupling, since there is not a similar Ward identity that relates the quark Yukawa coupling to the quark mass correction.

\section{$CP$-Violating $hW^+W^-$ Coupling in the Left-Right Model}\label{sec_LR}
\begin{figure}[!ht]
\centering
\includegraphics[width = 0.75 \linewidth]{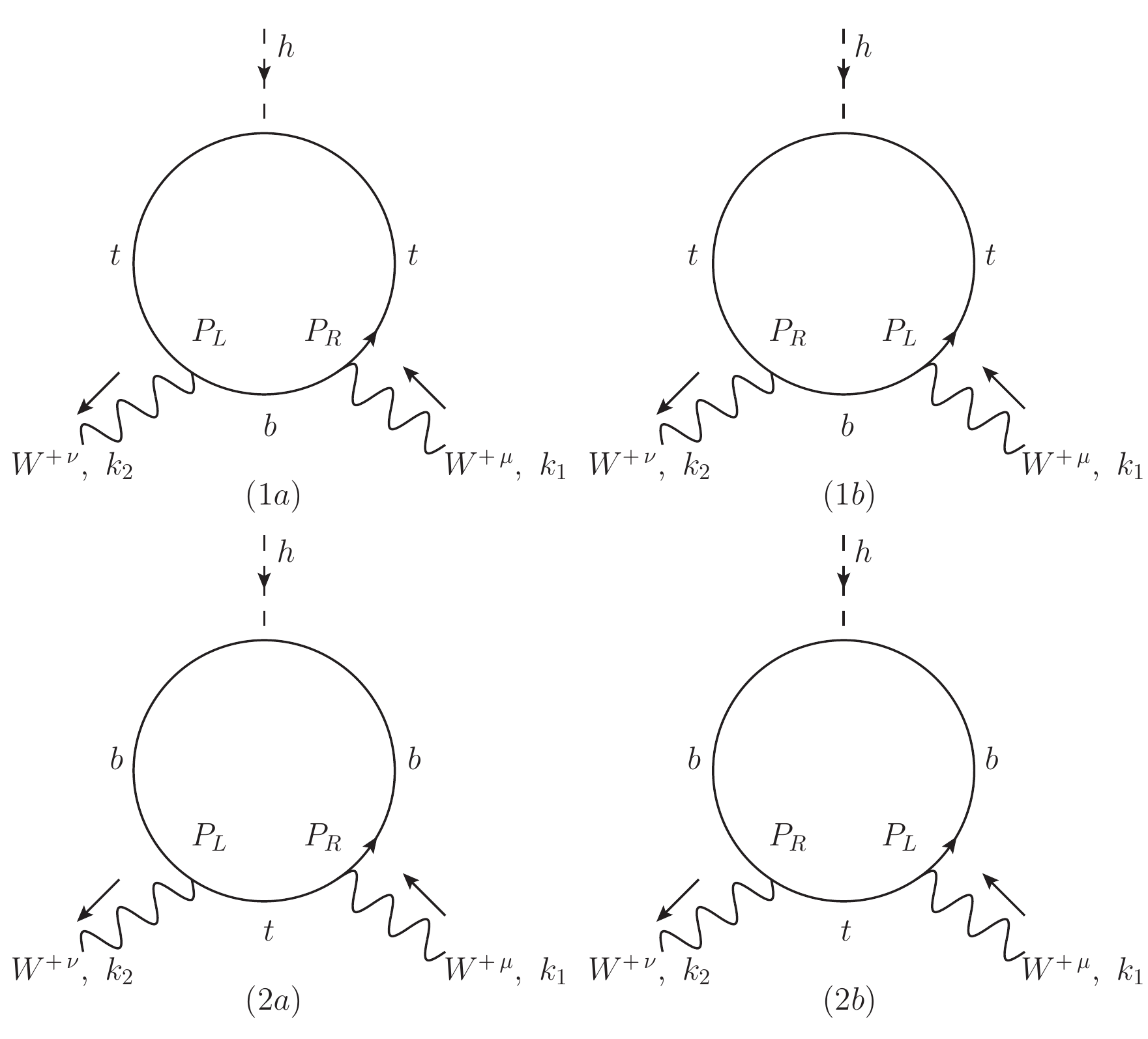}
\caption{Feynman diagrams to generate the CPV $hW^+W^-$ coupling in the left-right model.  }\label{FigLR}
\end{figure}

We now discuss the left-right model as proposed in~\cite{Pati:1974yy, Mohapatra:1974gc, Senjanovic:1975rk, Fritzsch:1975yn}. In this model one introduces a heavy $W$-boson for the right-handed gauge $SU(2)_R$ symmetry due to its inherent parity symmetry. After integrating over this heavy $W$-boson, the active light $W$-boson has the following general charged current
\begin{eqnarray}\label{Wcoupling}
{\cal L}^{\rm LR} &\supset & -\frac{g}{\sqrt{2}}W^+_\mu \sum_{i,j} \bar{u}_i\gamma^\mu (V_{u_i  d_j}P_L + U_{u_i d_j} P_R) d_j + {\rm h.c.}\,,
\end{eqnarray}
which arises from the left-right $W$-boson mixing. In this model a non-zero CPV coefficient appears at the one-loop level and even if only one generation of quarks is considered.
It is usually assumed that the CPV contribution is dominated by the third-generation quarks, $t$ and $b$ because the Feynman diagrams include Higgs couplings
to top and bottom quarks. Considering just the third generation the amplitude should be proportional to the factor ${\rm Im}(V_{tb} U^*_{tb}) \approx \zeta \sin\delta_{LR}$, where $\zeta$ stands for the mixing angle between the left and right $W$-boson and $\delta_{LR}$ denotes the phase related to the spontaneously $CP$ violation. Currently, the best upper limit on this factor is given by~\footnote{A more appropriate way to present the constraint in the left-right model is given by~\cite{Dekens:2014ina}
\begin{eqnarray}
\left|\frac{g_R}{g_L} \sin\zeta Im (V_L^{ud*} V_R^{ud} e^{i\delta_{LR}} )\right| \leq 4\times 10^{-6}\,,
\end{eqnarray}
where $g_{L,R}$ denote the left- and right-handed $W$-boson $SU(2)_{L,R}$ gauge couplings and $V_{L,R}$ the corresponding left- and right-handed CKM matrices. Here we assume that $g_L = g_R$ and $V_L = V_R$ due to the discrete parity $P$ and/or charge-conjugation $C$ symmetries imposed on the left-right model.}~\cite{Dekens:2014ina}
\begin{eqnarray}\label{ConstLR}
{\rm Im}(V_{tb} U^*_{tb}) \leq 4\times 10^{-6}\,,
\end{eqnarray}
which is obtained by applying naive dimensional analysis~\cite{Manohar:1983md,Weinberg:1989dx} on the constraint on the neutron electric dipole moment $d_n \leq 2.9\times 10^{-26} e$~cm~\cite{Baker:2006ts}. With the charge-current interactions of the active $W$-boson in Eq.~(\ref{Wcoupling}), there are four Feynman diagrams contributing to the CPV $hW^+W^-$ coupling at the one-loop order, which are shown in Fig.~\ref{FigLR}. In the remainder of this section we will compute all diagrams in order to extract the CPV $hW^+W^-$ vertex. 

Let us begin by focusing on the diagrams (1a) and (1b), with amplitudes
\begin{eqnarray}\label{ampLR1a0}
i {\cal M}^{\rm LR}_{(1a)} &=& (-1) N_c \int_l {\rm Tr}\Bigg[\left(-\frac{ig}{\sqrt{2}}U_{tb}\gamma_\mu P_R\right)\frac{i}{\slashed{l}-m_b} \left(-\frac{ig}{\sqrt{2}}V^*_{tb}\gamma_\nu P_L\right)\frac{i}{\slashed{l} + \slashed{k}_2-m_t} \nonumber\\
&& \times \left(-\frac{iy_t}{\sqrt{2}}\right)\frac{i}{\slashed{l}+\slashed{k}_1 - m_t} \Bigg] \nonumber\\
&=& - \frac{N_c g^2 m_t m_b}{2v} (U_{tb}V^*_{tb}) \int_l \frac{{\rm Tr}\{\gamma_\mu \gamma_\nu [(\slashed{l}+\slashed{k}_2)(\slashed{l}+\slashed{k}_1)+m_t^2]P_L \}}{(l^2 -m_b^2)[(l+k_2)^2-m_t^2][(l+k_1)^2 - m_t^2]}\,,
\end{eqnarray}  

\begin{eqnarray}\label{ampLR1b0}
i {\cal M}^{\rm LR}_{(1b)} &=& (-1) N_c \int_l {\rm Tr}\Bigg[\left(-\frac{ig}{\sqrt{2}}V_{tb}\gamma_\mu P_L\right)\frac{i}{\slashed{l}-m_b} \left(-\frac{ig}{\sqrt{2}}U^*_{tb}\gamma_\nu P_R\right)\frac{i}{\slashed{l} + \slashed{k}_2-m_t} \nonumber\\
&& \times \left(-\frac{iy_t}{\sqrt{2}}\right)\frac{i}{\slashed{l}+\slashed{k}_1 - m_t} \Bigg] \nonumber\\
&=& - \frac{N_c g^2 m_t m_b}{2v} (V_{tb}U^*_{tb}) \int_l \frac{{\rm Tr}\{\gamma_\mu \gamma_\nu [(\slashed{l}+\slashed{k}_2)(\slashed{l}+\slashed{k}_1)+m_t^2]P_R \}}{(l^2 -m_b^2)[(l+k_2)^2-m_t^2][(l+k_1)^2 - m_t^2]}\,.
\end{eqnarray}  

Now we sum up the above two terms and extract the CPV part of the final expression, which is given by
\begin{eqnarray}\label{ampLR1m}
i{\cal M}^{\rm LR}_{1} &\equiv & i{\cal M}_{(1a)}^{\rm LR} + i{\cal M}_{(1b)}^{\rm LR} \nonumber\\
&\sim & -\frac{i N_c g^2 }{2} \frac{m_t m_b}{v} {\rm Im}(V_{tb}U_{tb}^*) \int_l \frac{{\rm Tr}\{\gamma_\mu \gamma_\nu [(\slashed{l}+\slashed{k}_2)(\slashed{l}+\slashed{k}_1)+m_t^2]\gamma_5 \}}{(l^2 -m_b^2)[(l+k_2)^2-m_t^2][(l+k_1)^2 - m_t^2]}\,,
\end{eqnarray}
where we have used the relation ${\rm Im}(V_{tb} U_{tb}^*) = -{\rm Im}(U_{tb}V_{tb}^*)$. Since the operator we are interested in is shown in Eq.~(\ref{EffectiveOperator}), we only need to focus on the CPV part of the above loop integral, which is proportional to $\epsilon_{\mu\nu\rho\sigma} k_1^{\rho} k_2^\sigma$ and can be obtained by taking the trace of the $\gamma$-matrices. By performing the loop integrals for the obtained CPV part with the usual Feynman parametrization, we get the final expression for the CPV $hW^+W^-$ amplitude
\begin{eqnarray}\label{ampLR1f}
i{\cal M}_1^{\rm LR} \sim \frac{iN_c g^2}{8\pi^2 v} \frac{m_t m_b}{m_W^2} {\rm Im}(V_{tb}U_{tb}^*) \epsilon_{\mu\nu\rho\sigma} k_1^\rho k_2^\sigma {\cal I}\left(\frac{m_t^2}{m_W^2}, \frac{m_b^2}{m_W^2}\right)\,,
\end{eqnarray}
where ${\cal I}(x,y)$ represents the final Feynman parameter integration defined as 
\begin{eqnarray}
{\cal I} (x,y) \equiv \int^1_0 d\alpha \frac{\alpha(1-\alpha)}{\alpha x +(1-\alpha)y-\alpha(1-\alpha)}\,.
\end{eqnarray}
Note that when deriving Eq.~(\ref{ampLR1f}), we assume the two external $W$-boson momenta to be on-shell, {\it i.e.}, $k_1^2 = k_2^2 = m_W^2$ and take the zero Higgs momentum limit $k_1 - k_2 \to 0$. 

For the remaining two Feynman diagrams (2a) and (2b) in Fig.~\ref{FigLR}, we can repeat the procedure for diagrams (1a) and (1b), yielding the following CPV amplitude
\begin{eqnarray}\label{ampLR2f}
i{\cal M}_2^{\rm LR} \sim \frac{iN_c g^2}{8\pi^2 v} \frac{m_t m_b}{m_W^2} {\rm Im}(V_{tb}U_{tb}^*) \epsilon_{\mu\nu\rho\sigma} k_1^\rho k_2^\sigma {\cal I}\left(\frac{m_b^2}{m_W^2}, \frac{m_t^2}{m_W^2}\right)\,. 
\end{eqnarray}
Now we note that the function ${\cal I}(x,y)$ is symmetric under the exchange of the two variables $x$ and $y$. Thus, we can sum Eqs.~(\ref{ampLR1f}) and (\ref{ampLR2f}) to obtain the final CPV $hW^+W^-$ amplitude 
\begin{eqnarray}\label{ampLRf}
i{\cal M}^{\rm LR} \sim \frac{iN_c g^2}{4\pi^2 v} \frac{m_t m_b}{m_W^2} {\rm Im}(V_{tb}U_{tb}^*)  {\cal I}\left(\frac{m_t^2}{m_W^2}, \frac{m_b^2}{m_W^2}\right) \epsilon_{\mu\nu\rho\sigma} k_1^\rho k_2^\sigma \,. 
\end{eqnarray}
By comparing with the CPV effective operator in Eq.~(\ref{EffectiveOperator}), it is easy to show that the leading-order formula of the corresponding Wilson coefficient in the left-right model is
\begin{eqnarray}
c_{\rm CPV}^{\rm LR} \approx \frac{N_c g^2}{8\pi^2} \frac{m_t m_b}{m_W^2} {\cal I}\left(\frac{m_t^2}{m_W^2}, \frac{m_b^2}{m_W^2}\right) \zeta \sin\delta_{LR}\,,
\end{eqnarray}
where we have used the relation ${\rm Im}(V_{tb} U_{tb}^*) = \zeta \sin\delta_{LR}$. By taking the largest allowed value of ${\rm Im}(V_{tb}U^*_{tb})$ which is constrained by Eq.~(\ref{ConstLR}), the Wilson coefficient in the left-right model can be estimated to be
\begin{eqnarray}
c_{\rm CPV}^{\rm LR} \simeq 9.1\times 10^{-10} \sim {\cal O}(10^{-9})\,.
\end{eqnarray}
Even though numerically the CPV effect arising from the left-right model is still too small to be probed in the near future, it is already much larger than that in the SM. Therefore, it is more promising to test the CPV $hW^+ W^-$ vertex in the left-right model.

\section{$CP$-Violating $hW^+ W^-$ Coupling in the Complex 2-Higgs-Doublet Model}\label{sec_C2HDM}
\begin{figure}[!ht]
\centering
\includegraphics[width = 0.3 \linewidth]{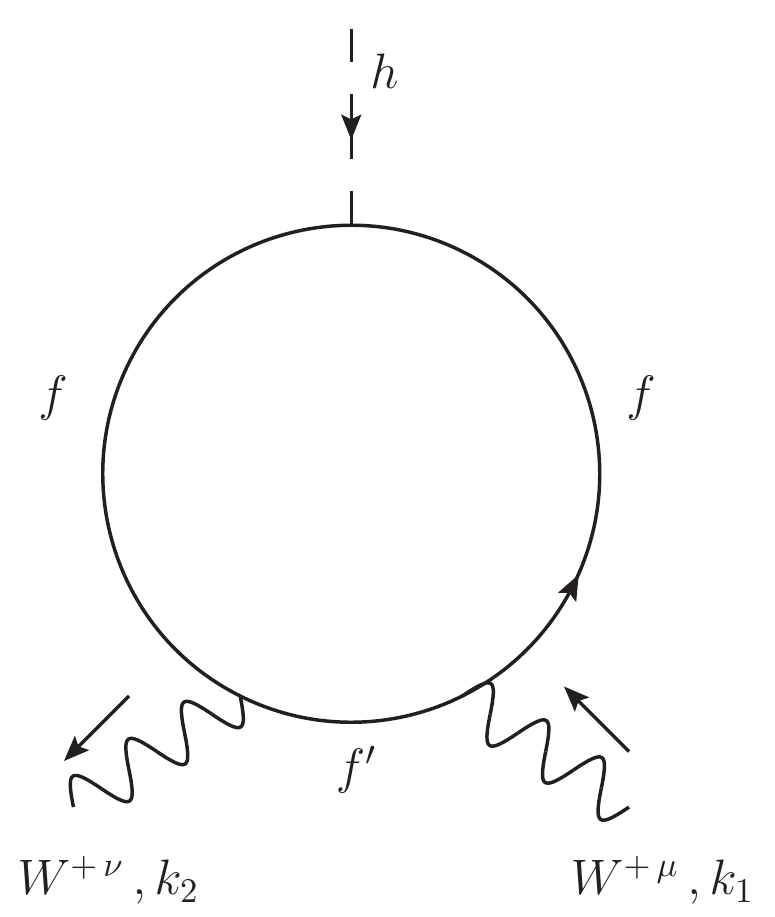}
\caption{An illustrative Feynman diagram to generate the CPV $hW^+W^-$ coupling in the C2HDM.  }\label{FigHD}
\end{figure}

The complex C2HDM~\cite{Lee:1973iz, Weinberg:1976hu,Ginzburg:2002wt} is one of the most popular models in which a new CPV source is generated from the scalar potential. In the present work, we focus on the computation of the effective CPV $hW^+ W^-$ vertex in the Type-II C2HDM as a simple CPV extension of the SM scalar sector~\cite{Fontes:2017zfn}. The model is built with two Higgs doublets instead of one together with an additional $Z_2$ symmetry. With this symmetry  problematic tree-level flavor-changing-neutral currents~\cite{Glashow:1976nt,Paschos:1976ay} are avoided and the model becomes simpler.The $Z_2$ symmetry is softly broken by a mass term which not only allows for the model to have a decoupling limit but it also introduces a unique CPV source in the scalar potential. After electroweak gauge symmetry breaking, the two doublets obtain their VEVs. The CPV source in the scalar sector induces mixing between all three neutral states and there are no
states with definite CP. As a result, the observed Higgs boson $h$ with mass $m_h = 125$~GeV is one of the mass eigenstates, and its Yukawa couplings to the SM fermions $\psi_f$ is modified as follows
\begin{eqnarray}\label{YukawaHD}
{\cal L}^{\rm C2HDM}_Y \supset -\sum_f \frac{m_f}{v} \bar{\psi}_f (c^e_f + i c_f^o \gamma_5)\psi_f h \,.
\end{eqnarray}  
It turns out that the leading-order contributions to the anomalous CPV $hW^+ W^-$ vertex can be generated at one-loop level as shown in Fig.~\ref{FigHD}. 
Note that the Yukawa couplings in ${\cal L}_Y^{\rm C2HDM}$ are proportional to the corresponding SM fermion masses, which dictates that the one-loop contribution is dominated by the diagram in which $f$ and $f^\prime$ in Fig.~\ref{FigHD} are identified to be the third-generation quarks, {\it i.e.}, $f,f^\prime = t, b$. 

Now let us begin by computing the one-loop diagram with $f=t$ and $f^\prime = b$, with the corresponding amplitude given by
\begin{eqnarray}
i{\cal M}^{\rm C2HDM}_{tb} &=& (-1) N_c \int_l {\rm Tr}\Bigg[ \left(-\frac{ig}{\sqrt{2}} V_{tb}\gamma_\mu P_L\right) \frac{i}{\slashed{l}-m_b} \left(-\frac{ig}{\sqrt{2}} V_{tb}^* \gamma_\nu P_L\right) \frac{i}{\slashed{l}+\slashed{k}_2-m_t} \nonumber\\
&& \times \left(-i \frac{m_t}{v}\right)(c_t^e + i c_t^o \gamma_5) \frac{i}{\slashed{l}+\slashed{k}_1-m_t}\Bigg] \nonumber\\
&=& -\frac{N_c g^2 m_t |V_{tb}|^2}{2v} \frac{{\rm Tr}[\gamma_\mu\slashed{l}\gamma_\nu P_L (\slashed{l}+\slashed{k}_2 + m_t)(c_t^e + i c_t^o \gamma_5)(\slashed{l}+\slashed{k}_1+m_t)]}{(l^2-m_b^2)[(l+k_2)^2-m_t^2][(l+k_1)^2-m_t^2]}\,.
\end{eqnarray}
By using the Feynman parameters and performing the loop integration over $l$, we can pick up the $P$-odd and $CP$-odd term in the above amplitude as follows:
\begin{eqnarray}\label{ampHD}
i{\cal M}^{\rm C2HDM}_{tb}\sim \frac{i g^2 N_c c_t^o}{16\pi^2 v} \frac{m_t^2}{m_W^2} |V_{tb}|^2 \epsilon_{\mu\nu\rho\sigma} k_1^\rho k_2^\sigma {\cal I}_{1} \left(\frac{m_t^2}{m_W^2}, \frac{m_b^2}{m_W^2}\right)\,,
\end{eqnarray}
where the remaining Feynman parameter integration ${\cal I}_1 (x,y)$ is given by
\begin{eqnarray}
{\cal I}_1 (x,y) \equiv \int^1_0 d\alpha \frac{\alpha^2}{\alpha x + (1-\alpha) y-\alpha(1-\alpha)}\,.
\end{eqnarray}
Note that the CPV amplitude in Eq.~(\ref{ampHD}) is proportional to $m_t^2$. Thus, if the $CP$-odd Yukawa coupling are of the same order for all quark and lepton flavors, this amplitude would usually dominate the induced CPV $hW^+W^-$ vertex. However, in the Type-II C2HDM~\cite{Fontes:2017zfn}, the $CP$-odd couplings of the up-type quarks are inversely proportional to the quantity ${\rm tan}\beta$, while those for the down-type quarks are proportional to ${\rm tan}\beta$, where ${\rm tan}\beta \equiv v_2/v_1$ characterises the ratio between the vacuum expectation values $v_{1,2}$ of the two Higgs doublets. When ${\rm tan}\beta$ becomes large, of the order of $m_t^2/m_b^2$,  the $CP$-odd top-quark coupling becomes comparable to the bottom one, and the diagram with $f=b$ and $f^\prime = t$ 
as to be considered for the CPV $hW^+W^-$ effect, with the amplitude given by
\begin{eqnarray}\label{ampHD1}
i{\cal M}^{\rm C2HDM}_{bt}\sim \frac{i g^2 N_c c_b^o}{16\pi^2 v} \frac{m_b^2}{m_W^2} |V_{tb}|^2 \epsilon_{\mu\nu\rho\sigma} k_1^\rho k_2^\sigma {\cal I}_{1} \left(\frac{m_b^2}{m_W^2}, \frac{m_t^2}{m_W^2}\right)\,.
\end{eqnarray}
In sum, by comparing with the effective operator in Eq.~(\ref{EffectiveOperator}), the general dominant Wilson coefficient can be obtained as follows:
\begin{eqnarray}
c_{\rm CPV}^{\rm C2HDM} = \frac{N_c g^2}{32\pi^2} |V_{tb}|^2 \Bigg[\frac{c_t^o m_t^2}{m_W^2}{\cal I}_1 \left(\frac{m_t^2}{m_W^2}, \frac{m_b^2}{m_W^2}\right) + \frac{c_b^o m_b^2}{m_W^2}{\cal I}_1 \left(\frac{m_b^2}{m_W^2}, \frac{m_t^2}{m_W^2}\right) \Bigg]\,.
\end{eqnarray}       

In order to give an estimate of the CPV Wilson coefficient in the C2HDM, we can apply the latest fitting results on the $CP$-odd quark Yakawa coupling $c_f^o$ in Ref.~\cite{Fontes:2017zfn}. In particular, if the observed Higgs $h$ is the lightest neutral scalar in the spectrum, $c_t^o$ can be as large as 0.3. In this case, the diagram with $f=t$ and $f^\prime = b$ is expected to dominate the CPV amplitude, with the size of the Wilson coefficient estimated to be
\begin{eqnarray}
c_{\rm CPV}^{\rm C2HDM} \simeq 6.6\times 10^{-4} \sim {\cal O}(10^{-3})\,.
\end{eqnarray}  

In the light of such a large CPV $hW^+W^-$ coupling in the C2HDM that might be measured in the future collider experiments, we calculate the related CPV $hZZ$ coupling in the C2HDM in the Appendix~\ref{sec_hZZ}. 

\section{Conclusion}\label{sec_Conclusion}

There is a great effort in the community in the study of the properties of the observed Higgs boson at the LHC. As CP violation in the scalar sector is a major issue 
at the LHC and future colliders, it is essential to use all observables at hand to understand the properties of the Higgs boson. One important quantity is the anomalous CPV $hW^+ W^-$ coupling, 
which can be represented by either the scattering amplitude in Eq.~(\ref{amp0}) or the effective operator in Eq.~(\ref{EffectiveOperator}). 
In the light of the recent experimental developments, we studied the size of this CPV $hW^+ W^-$ effect in the SM and two BSM benchmark models: the left-right model and the C2HDM. In the SM, we found that the leading-order contribution arises at the two-loop level. Further, by the explicit summation over the up- and down-type quark flavors in the loop integrals, we have shown that the corresponding total amplitude or the Wilson coefficient should be proportional to the factor in Eq.~(\ref{CommFac}), which is actually the reflection of the GIM mechanism. Based on this observation, we have further estimated the order of the Wilson coefficient of the induced $hW^+ W^-$ operator to be approximately ${\cal O}(10^{-23})$, which is too small to  be observed under the present experimental technology. 
On the other hand, for the two benchmark models beyond the SM, the CPV $hW^+ W^-$ interaction can be much larger than that in the SM, partly due to the fact that this CPV phenomenon already exists at the one-loop level. As a result, the Wilson coefficient in the left-right model can be of the order ${\cal O}(10^{-9})$, while it can be further boosted to ${\cal O}(10^{-3})$ in the case of the C2HDM. The present predictions for the High-Luminosity LHC~\cite{CMS:2018qgz} and for a future International Linear Collider~\cite{Barklow:2015tja} with $\sqrt{s} = 500$ GeV are at the moment of the order ${\cal O}(10^{-2})$. Therefore models such as the C2HDM may be within the reach of these machines.

\appendix
\section{$CP$-Violating $hZZ$ Coupling in the C2HDM}\label{sec_hZZ}
\begin{figure}[!ht]
\centering
\includegraphics[width = 0.75 \linewidth]{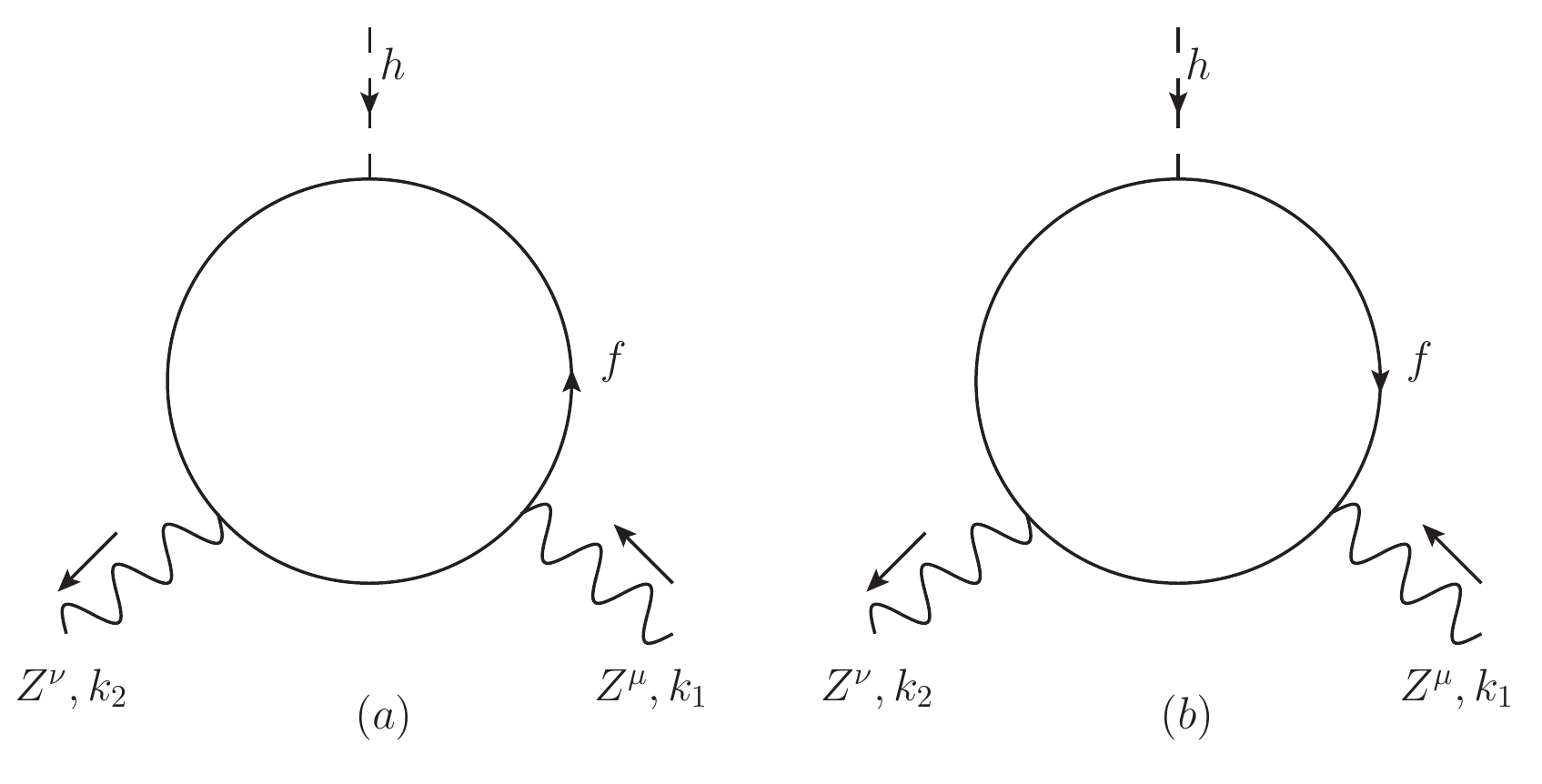}
\caption{Feynman diagrams generating the CPV $hZZ$ coupling in the C2HDM.  }\label{FigZZ}
\end{figure}

In face of the possible importance of the CPV $hW^+W^-$ coupling in the C2HDM, we will extend the discussion to the CPV $hZZ$ coupling, which can be represented by the following effective operator
\begin{eqnarray}\label{OZZ}
{\cal O}^{ZZ}_{\rm CPV} = - \frac{c_{\rm CPV}^{ZZ}}{v} h Z^{\mu\nu} \tilde{Z}_{\mu\nu}\,,
\end{eqnarray}
where $Z^{\mu\nu}$ and $\tilde{Z}_{\mu\nu}$ denote the $Z$-boson field strength and its dual. Our task in this section is to compute the leading-order contribution to this effective operator in the C2HDM, which, like its $hW^+W^-$ counterpart, should be induced at one-loop level by the Feynman diagrams shown in Fig.~\ref{FigZZ}. In order to proceed, we note that the $Z$-boson coupling to any SM fermion $f$ can be written as
\begin{eqnarray}
\left(-i\frac{2m_Z}{v}\gamma^\mu\right) (T_{3f} P_L -Q_f s_W^2)\,,
\end{eqnarray} 
where $m_Z$ is the $Z$-boson mass, $s_W \equiv \sin\theta_W$ with $\theta_W$ the Weinberg angle, and $T_{3f}$ ($Q_f$) is the isospin (electric) charge of the fermion $f$. With this notation, we can write down the amplitudes of both diagrams as
\begin{eqnarray}
i{\cal M}^{\rm C2HDM}_{ZZ (a)} =& & (-N_c) \int_l {\rm Tr} \Bigg[\left(-i \frac{2m_Z}{v} \gamma_\mu \right) (T_{3f}P_L - Q_f s_W^2) \frac{i}{\slashed{l}-m_f} \nonumber\\
&& \left(-i \frac{2m_Z}{v} \gamma_\nu \right) (T_{3f}P_L - Q_f s_W^2) \frac{i}{\slashed{l}+\slashed{k}_2-m_f} \nonumber\\
&& \left(-i \frac{m_f}{v} \right) (c_f^e + i c_f^o \gamma^5)  \frac{i}{\slashed{l}+\slashed{k}_1-m_f}\Bigg]\,,
\end{eqnarray} 
\begin{eqnarray}
i{\cal M}^{\rm C2HDM}_{ZZ (b)} =& & (-N_c) \int_l {\rm Tr} \Bigg[\left(-i \frac{2m_Z}{v} \gamma_\nu \right) (T_{3f}P_L - Q_f s_W^2) \frac{i}{\slashed{l}-m_f} \nonumber\\
&& \left(-i \frac{2m_Z}{v} \gamma_\mu \right) (T_{3f}P_L - Q_f s_W^2) \frac{i}{\slashed{l}-\slashed{k}_1-m_f} \nonumber\\
&& \left(-i \frac{m_f}{v} \right) (c_f^e + i c_f^o \gamma^5)  \frac{i}{\slashed{l}-\slashed{k}_2-m_f}\Bigg]\,,
\end{eqnarray} 
where the external momentum $k_1$ flows into the loop while $k_2$ flows out. 

By integrating over the loop momentum $l$, we find that both diagrams give exactly the same contribution to the CPV $hZZ$ coupling. Therefore, the total CPV $hZZ$ amplitude is given by
\begin{eqnarray}
i{\cal M}^{\rm C2HDM}_{ZZ} \sim \frac{i N_c c_f^o}{\pi^2} \frac{m_f^2}{v^3}  \epsilon_{\mu\nu\rho\sigma} k_1^\rho k_2^\sigma \int^1_0 dt \frac{t(2Q_f^2 s_W^4 - 2Q_f s_W^2 T_{3f}+T_{3f}^2 t)}{m_f^2/m_Z^2 - t(1-t)}\,,
\end{eqnarray}  
where we assume the external $Z$-boson momenta to be on-shell with $k_1^2 = k_2^2 = m_Z^2$ and take the limit of the vanishing Higgs momentum squared. Therefore, by matching the effective operator in Eq.~(\ref{OZZ}), the Wilson coefficient is given by
\begin{eqnarray}
c_{\rm CPV}^{ZZ\,{\rm C2HDM}} = \frac{N_c c_f^o}{4\pi^2} \frac{m_f^2}{v^2} \int^1_0 dt \frac{t(2Q_f^2 s_W^4 - 2Q_f s_W^2 T_{3f}+T_{3f}^2 t)}{m_f^2/m_Z^2 - t(1-t)}\,.
\end{eqnarray}
Considering just the top quark contribution, we can estimate the size of this CPV $hZZ$ effective, with the corresponding value of $c_{\rm CPV}^{ZZ}$ is given by
\begin{eqnarray}
c_{\rm CPV}^{ZZ\,{\rm C2HDM}} \simeq {\cal O}(10^{-4})\,,
\end{eqnarray}
where we have used $c_t^o = 0.3$ that is the largest value allowed by the current experimental constraints~\cite{Fontes:2017zfn}.

\section*{Acknowledgments}
DH and APM are supported by the Center for Research and Development in Mathematics and Applications (CIDMA) through the Portuguese Foundation for Science and Technology (FCT - Fundação para a Ciência e a Tecnologia), references UIDB/04106/2020 and UIDP/04106/2020. DH, APM and RS are supported by the project PTDC/FIS-PAR/31000/2017. DH is also supported by the Chinese Academy of Sciences (CAS) Hundred-Talent Program. RS is also supported by FCT, Contracts UIDB/00618/2020, UIDP/00618/2020, CERN/FISPAR/0002/2017, CERN/FIS-PAR/0014/2019, and by the HARMONIA project, contract UMO-2015/18/M/ST2/0518. APM is also supported by the projects CERN/FIS-PAR/0027/2019 and CERN/FISPAR/0002/2017 as well as by national funds (OE), through FCT, I.P., in the scope of the framework contract foreseen in the numbers 4, 5 and 6 of the article 23, of the Decree-Law 57/2016, of August 29, changed by Law 57/2017, of July 19.


\bibliography{CPV1}
\bibliographystyle{jhep}

\end{document}